# Hydex waveguides for nonlinear optics

D. J. Moss[1,4], A. Pasquazi[2], M. Peccianti[2], L Razzari[2], D. Duchesne[2], M. Ferrera[2], S. Chu[3], B.E. Little[3] and R. Morandotti[2]

[1]*CUDOS, School of Physics, University of Sydney, New South Wales 2006 Australia*
[2] *INRS-EMT, 1650 Boulevard Lionel Boulet, Varennes, Québec, Canada, J3X 1S2*
[3]*Infinera Corp, 9020 Junction Dr, Annapolis, Maryland, USA 94089*
[4]*Present Address: School of Electrical and Computer Engineering, RMIT University, Melbourne, Vic. Australia 3001*

**Abstract:** We demonstrate a wide range of novel functions in integrated, CMOS compatible, devices. This platform has promise for telecommunications and on-chip WDM optical interconnects for computing.

1. **Introduction**

All-optical signal processing has been demonstrated extensively in silicon including demultiplexing at 160Gb/s via four-wave mixing (FWM) [1] and optical regeneration in silicon nanowires [2], as well as optical regeneration and demultiplexing [3,4,5] in chalcogenide glass (ChG) waveguides. The third order nonlinear efficiency of all-optical devices can be improved by increasing the waveguide nonlinear parameter, $\gamma = \omega\, n_2 / c\, A_{eff}$ (where $A_{eff}$ is the effective area of the waveguide, $n_2$ is the Kerr nonlinearity, and $\omega$ is the pump frequency) as well as by using resonant structures to enhance the local field intensity. High index materials, such as semiconductors and ChG, offer excellent optical confinement and high values of $n_2$, a powerful combination that has produced extremely high values of $\gamma$: 200,000 $W^{-1}$ $km^{-1}$ for silicon nanowires [2], and 93,400 $W^{-1}$ $km^{-1}$ in ChG nanotapers [6]. Yet silicon suffers from high nonlinear losses due to two-photon absorption (TPA) and the resulting free carriers [7]. Even if free carriers can be eliminated by using p-i-n junctions, its poor intrinsic nonlinear figure of merit (FOM = $n_2 / (\beta\, \lambda)$, where $\beta$ is the two-photon absorption coefficient) is very low. While this FOM is considerably higher for ChG, the development of fabrication processes for these newer materials is at a much earlier stage.

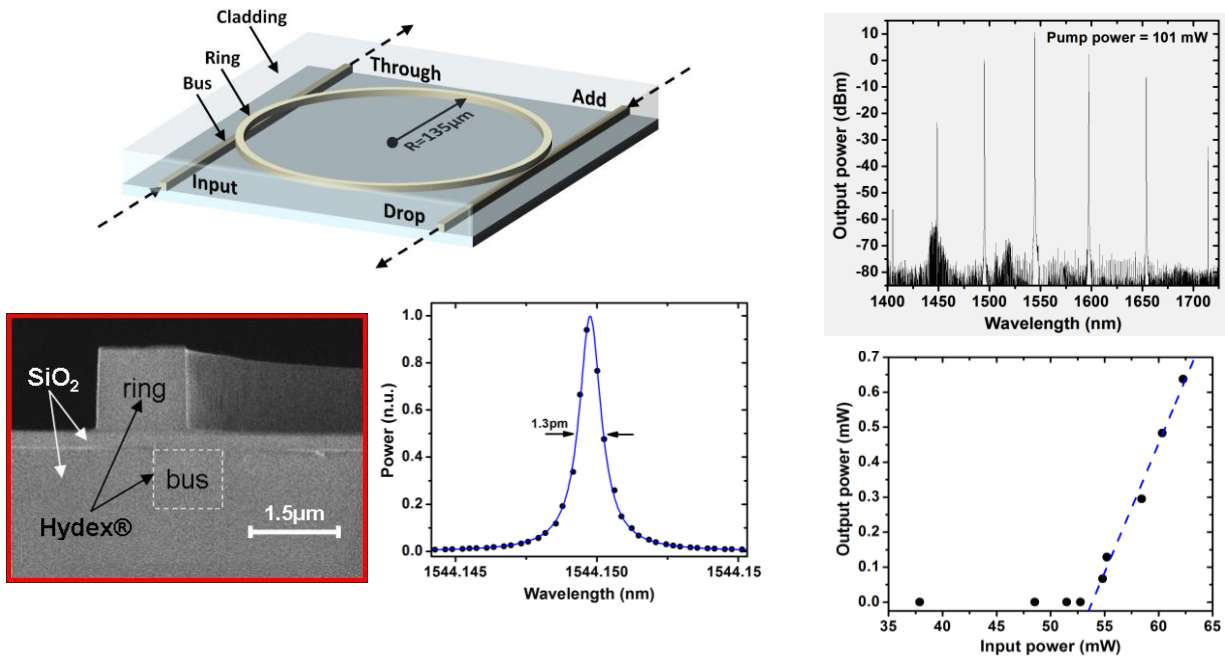

Figure 1. Top Left: Device schematic - 4 port micro-ring resonator (fiber pigtails not shown), Bottom Left: SEM cross-section before depositing upper cladding of SiO, Bottom Middle: linear optical transmission spectra showing a 1.3pm FWHM. Top Right: Output spectra in the drop waveguide of the hyperparametric oscillator when pumping in the anomalous dispersion regime at 1544.15nm and Bottom Right: output in the drop waveguide for the oscillating mode at 1596.98nm vs pump power (at 1544.15nm) showing a slope efficiency of 7.4% above threshold.

Here, we report a wide range of functions in a CMOS compatible high index doped silica glass platform, including a efficient four wave mixing [8], an integrated multiple wavelength source [9], ultra-high repetition rate modelocked laser [10], and all-optical integrator [11] based on integrated ring resonators. We achieve CW optical "hyper-parametric" oscillation in a high index, high Q factor (1.2 million) silica glass micro-ring resonator with a differential slope efficiency above threshold of 7.4% for a single oscillating mode out of a single port, a CW threshold power as low as 54mW, and a controllable range of frequency spacing from 200GHz to more than 6THz. We also achieve modelocking in this device with pulse repetition rates as high as 800GHz [10]. Finally, we also report novel functions in ultra-long (45cm) spiral waveguides, including a time-lens based optical oscilloscope [12] and integrated parametric amplifier with > 15dB of gain [13] and an integrated pulse compressor [14]. The success of these devices was due to very low linear loss, a high nonlinearity parameter $\gamma = \omega \, n_2 \,/\, (c \, A_{eff}) \cong 233 W^{-1} km^{-1}$ (where $A_{eff}$ is the effective waveguide area), as well as to negligible nonlinear losses up to extremely high intensities (25GW/cm$^2$) [15]. The low loss, design flexibility, and CMOS compatibility of this device will enable multiple wavelength sources for telecommunications, computing, sensing, metrology and other areas.

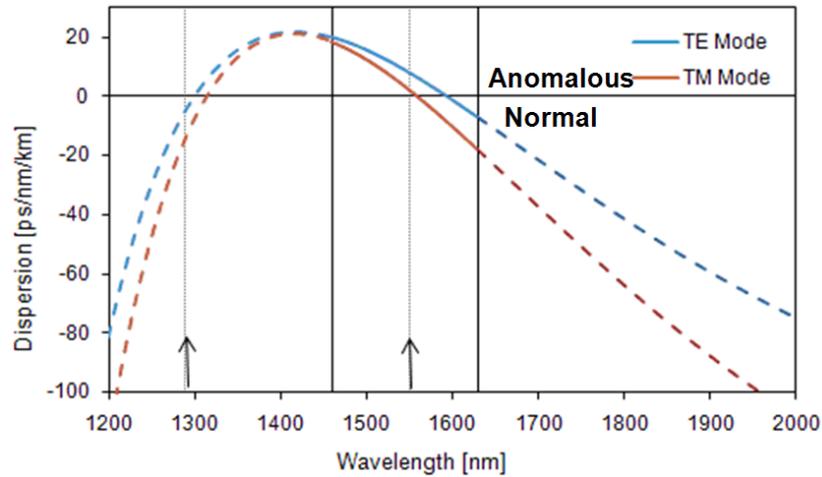

Figure 2. Waveguide dispersion for TE and TM polarizations. The solid lines represent experimental measurements while the dashed lines represent extrapolations based on modeling. The vertical arrows denote the zero dispersion points in the waveguide.

2. **Ring Resonator Devices**

Figure 1 shows the device structure for the integrated hyper-parametric oscillator - a four port micro-ring resonator with radius $\cong$ 135μm, along with an SEM of the waveguide (1.45μm x 1.5μm). The bus waveguides used to couple light in and out of the resonator have the same cross section and are buried in $SiO_2$. The waveguide core is low loss, high index (n=1.7) doped silica glass with a core-cladding contrast of 17% [9]. The films were deposited by standard chemical vapor deposition (CVD) and device patterning and fabrication were performed by photolithography and reactive ion etching before over-coating with silica glass. Propagation losses were 0.04dB/cm and coupling loss to fiber pigtails of $\cong$ 1.5dB / facet. Figure 1 shows the linear transmission spectra of the resonator from the input to the drop port, with a free spectral range (FSR) of 200GHz and a FWHM bandwidth of 1.3pm, corresponding to a Q factor of 1.2 million. The dispersion in these waveguides is extremely low and anomalous [16] over most of the C-band (Figure 2), which is nearly ideal for FWM.

Figure 1 shows the output spectra for a 101mW (inside the waveguide) TM polarized pump tuned to the resonance at 1544.15nm showing lasing at 1596.98nm - 52.83nm away from the pump. As the power is increased above threshold (53.8mW) both the wavelength of the mode and the pump power in the cavity are clamped as cascaded FWM takes over, generating further wavelengths with equal spacing. For a pump at 1544.15nm the theoretical MI gain curve peaks at ~1590nm, which agrees well with the observed lasing wavelength of 1596.98nm. Figure 1 also shows the power of the mode at 1596.98nm exiting one port (drop) versus input pump power, showing a (single line) differential slope efficiency above threshold of 7.4%. The maximum total output power was at 101mW pump at 1544.15nm (Fig 1b), where we obtained 9mW in all modes out of both ports, with 2.6mW in a single line at 1596.98nm from a single port. The total oscillating mode power of 9mW represents a total conversion efficiency of 9%. The zero dispersion point for TM polarization is ~1560nm in these waveguides (Figure 2) with $\lambda < 1560$nm anomalous and $\lambda > 1560$nm normal. When pumping at 1565.19 nm (normal dispersion) we observed no oscillation, consistent with parametric gain due to FWM and MI. When pumping near zero dispersion, at 1558.65nm, we observed lasing with a spacing of 28.15nm, agreeing with the expected shift in the MI gain profile. We also achieve results for an integrated subpicosecond soliton laser [10], where we obtained pulse repetition rates up to 800GHz, and with pulse-widths well below 1ps. Next we turn to the results for devices based on a 45cm integrated spiral waveguide.

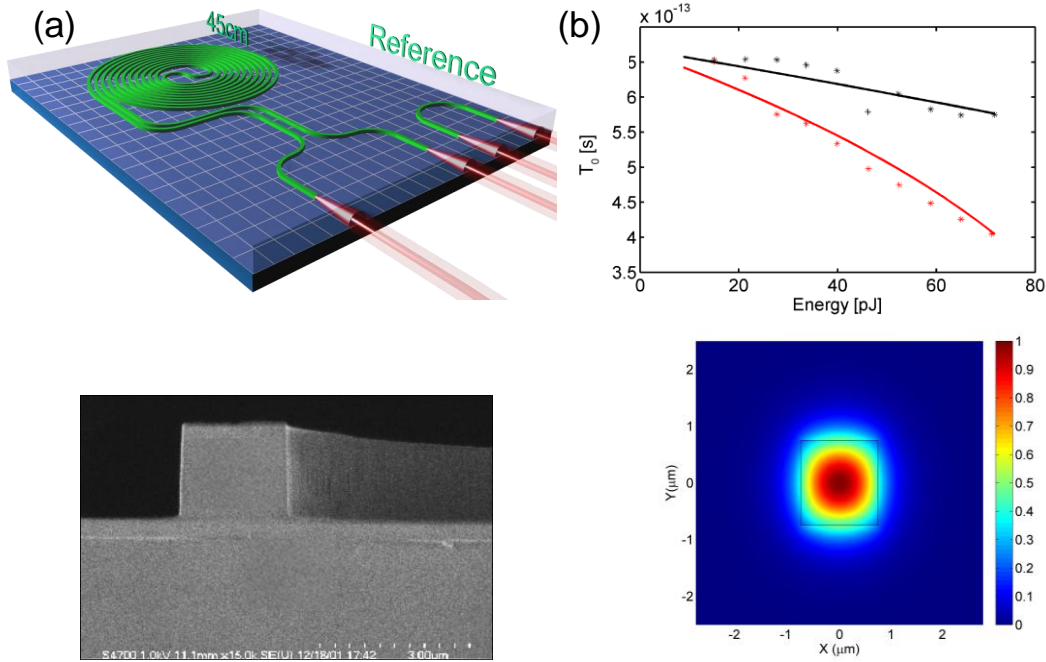

Figure 3. Top Left: Device schematic – 45cm long spiral waveguide, Bottom Left: SEM cross-section before depositing upper cladding of SiO, Bottom Right: Mode profile from modeling. Top Right: Results of pulse compression measurements.

3. **Spiral Waveguide Devices**

We report a wide range of functions in ultra-long (45cm) spiral waveguides in this same CMOS compatible high index doped silica glass platform [8], including optical pulse compression [14], a time-lens based optical oscilloscope [12] and integrated parametric amplifier with > 15dB of gain [13]. As in our work on the ring resonators, the success of these spiral waveguide devices is due to very low linear loss, a high nonlinearity parameter $\gamma = \omega n_2 / (c A_{eff}) \cong 233$W$^{-1}$km$^{-1}$ (where $A_{eff}$ is the effective waveguide area), as well as to negligible nonlinear losses up to extremely high intensities (25GW/cm$^2$) [15]. The low loss, design flexibility, and CMOS compatibility of this device will enable multiple wavelength sources for signal processing in telecommunications [17-22], computing, sensing, metrology and other areas.

Figure 3 shows the device structure for the integrated waveguide - a 45cm long spiral - with an SEM of the waveguide (1.45μm x 1.5μm) that is buried in SiO$_2$. The waveguide core is low loss, high index (n=1.7) doped silica glass with a core-cladding contrast of 17% [8]. The fabrication process was similar to that of the ring resonators. The films were deposited by standard chemical vapor deposition (CVD) and device patterning and fabrication were performed by photolithography and reactive ion etching before over-coating with silica glass. Propagation losses were 0.04dB/cm and coupling loss to fiber pigtails of $\cong$ 1.5dB / facet. The dispersion in these waveguides was the same as for the ring resonators - extremely low and anomalous (Figure 2) over most of the C-band - nearly ideal for FWM. The zero dispersion point for TM polarization is ~1560nm in these waveguides [16], with λ < 1560nm anomalous and λ > 1560nm normal.

Figure 3 also shows experimental results for optical pulse compression [14] via on-chip self-phase modulation followed by dispersion compensation together with theory, calculated by integrating the standard Nonlinear Schrödinger equation (NLSE), including the effects of the input (7.33m) and output (0.66m) fiber patch cords, which predominantly contribute only linear dispersion. Figure 3 shows good agreement between theory and experiment, as well as experimental results that show pulse compression from about 650fs down to 400fs, with about 50fs of this arising from dispersion accumulated in the fiber patch cords.

Figure 4 shows the results [13] for optical parametric gain using sub-picosecond pulses obtained from a Spectra Physics Tsunami Ti:Sappire 180fs 80MHz laser pumping an OPO. The output pulses (bandwidth = 30nm) were split and filtered with two tunable Gaussian filters of 5nm bandwidth (700fs), in order to obtain pump and signal pulses. The pump and signal pulses were combined into standard SMF using a 90/10% beam splitter and then coupled into the spiral. Pulse synchronization was adjusted with a delay line, while power and polarization were controlled with a polarizer and λ/2 plate. Figure 4 shows the results for a pump wavelength $\lambda_{pump}$=1525nm and probe (signal) wavelengths at $\lambda_{signal}$=1480; 1490; 1500nm. The pump peak power coupled inside the waveguides was varied from 3 to 40W, while the signal peak power was 3mW. The signal was efficiently converted to the idler wavelength in the C band with $\lambda_{idler}$=1578; 1565; 1547nm. For pump powers larger than 30W the idler is strong enough to induce cascaded FWM. A small asymmetry is visible in the interaction (the cascaded FWM induced by the idler is stronger than the one induced with the signal). This occurrence indicates a non negligible effect of $\beta_3$, consistent with a low absolute value of $\beta_2$. We define the conversion efficiency as the ratio of the transmitted pulse energy of the idler, to the transmitted signal without the pump [13] (so not including coupling losses), in order to account for spectral broadening due to XPM. The measured efficiency vs pump peak power (Figure 5, $\lambda_{signal}$=1480nm) shows a remarkable peak efficiency of +16.5dB at 40W pump power.

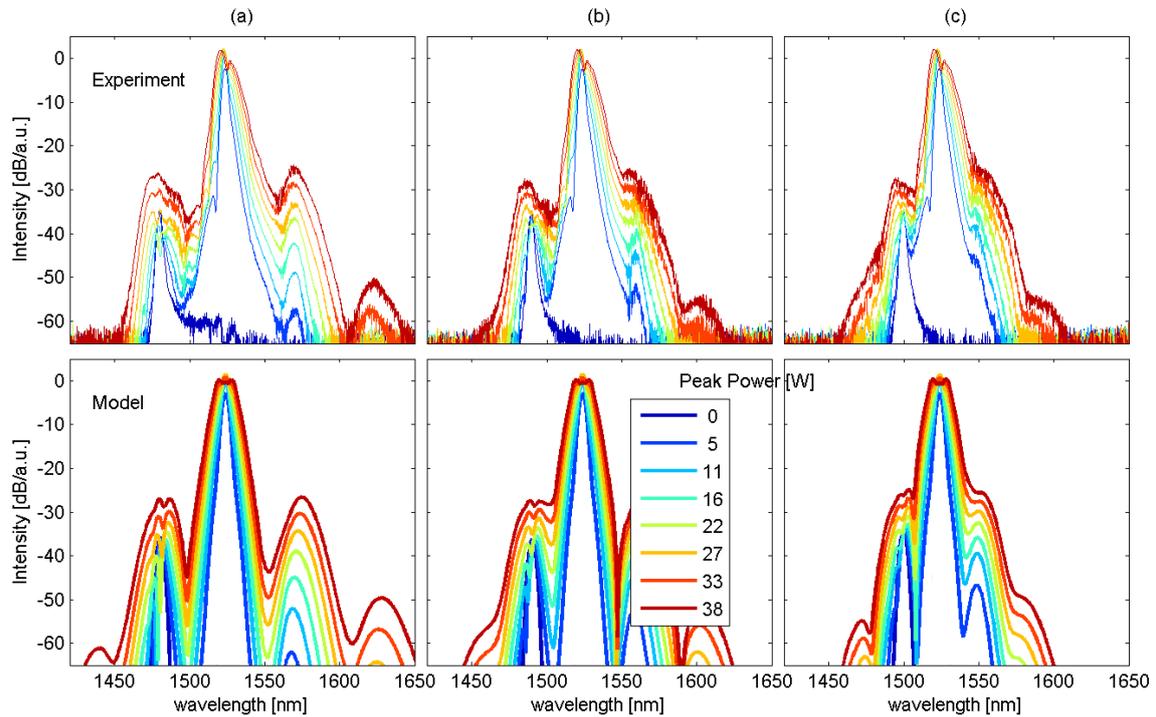

Figure 4. Experiments and (middle) theory of parametric gain for a 1525nm pump and 1480; 1490; 1500nm from respectively (a),(b),(c). The legend lists pump peak powers, while signal peak power is 3mW.

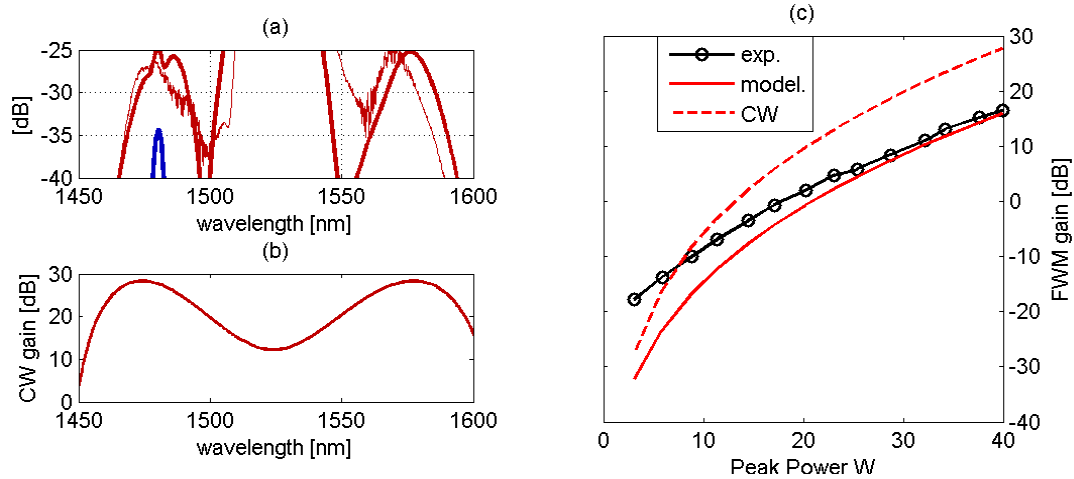

Figure 5. Gain for 1480nm signal: (a) Spectra for a 3mW peak power signal alone (blue) and with 40W pump (red). (b): CW gain. (c) FWM gain: measurement (black dots) model in the experimental conditions (red continuous line) and CW (dashed)

Next, we use this FWM process as a basis for time-lens temporal pulse imaging on a chip [12]. A sketch of the experimental setup for the time-lens measurements is shown in Figure 6. Pump and signal are obtained filtering a 1540nm, 30nm bandwidth OPO pulse at 80MHz rep rate. A Gaussian 8nm bandwidth tunable filter is used for the pump, while a 5 or 8nm bandwidth filter is used for the signal, thus obtaining two new peaks centered at 1530 and 1560nm respectively. An interferometer splits the signal in two pulses (signal under test), and a movable mirror controls the relative delay. The pulses are dispersed with a standard single mode fibre (SMF). The pump is dispersed twice as much as the signal length: hence the signal chirp is exactly compensated by the TL chirp in the FWM process and the idler results in the FT of the input signal. We kept the pulses power in the SFM low enough to avoid SPM. Pulses are first amplified with a standard erbium doped fiber amplifier providing 25mW total average power, and are then coupled in the 45nm waveguide. An optical spectrum analyzer provides the output spectrum. A typical result is shown in Figure 7. The SPM free pump is visible at 1530nm; the signal spectrum centered at 1560 shows clearly interference of the two delayed pulses. The FWM product around 1502nm is the temporal image of the signal under test. The system calibration is obtained by changing the delay between the two signal pulses, while the temporal window is measured by varying the pump-signal delay. We obtain a calibration factor of 1.77ps/nm and a -3dB window of 14ps. Figure 7 shows the measurement of a subpicosecond detail, namely the destructive interference between two replicas of a Gaussian pulse with 5nm bandwidth.

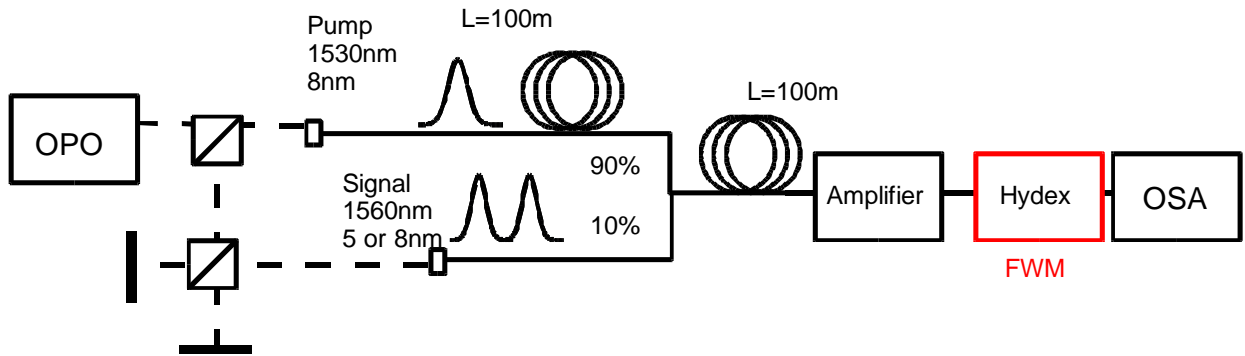

Figure 6. Experimental setup for time-lens measurements.

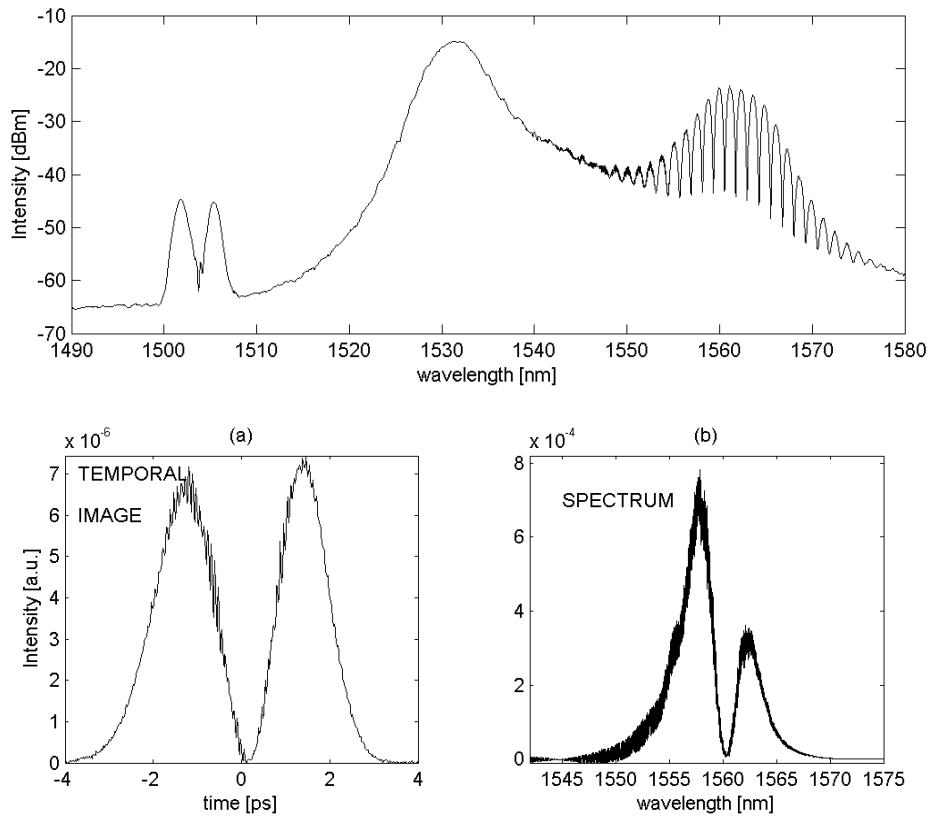

Figure 7. Top: OSA spectrum: The pump is centered at 1530 nm. The spectrum of the signal under test (consisting in two delayed pulses) is visible at 1560nm. The interference of the two replicas is evident. The FWM idler (around 1502nm) clearly shows an image of the signal under test (double pulse). Bottom: time (a) temporal image and (b) spectrum of two 5nm bandwidth Gaussian replicas showing destructive interference.

4. **Conclusions**

We demonstrate a wide range of functions in CMOS compatible high index doped silica glass devices, including low power CW FWM, hyper parametric oscillation, optical pulse compression, parametric gain, and time-lensed imaging. This platform has strong potential for photonic integrated circuits for telecommunications and on-chip WDM optical interconnects for computing.